\documentclass[runningheads]{svmult}

\usepackage{graphicx}  
\usepackage{multicol}  
\usepackage{physprbb}  

\newcommand{\be}{\begin{equation}}
\newcommand{\ee}{\end{equation}}
\newcommand{\hr}{\hat\varrho}
\newcommand{\ad}{\hat a{}^\dagger}

\begin{document}
\title*{Decoherence and Quantum-state Measurement in Quantum Optics}
\toctitle{Decoherence and Quantum-state Measurement in Quantum Optics}
%
%
\titlerunning{Decoherence and Quantum-state Measurement in Quantum Optics}
%
\author{Luiz Davidovich}
\authorrunning{Luiz Davidovich}
%
%
\institute{Instituto de F\'\i sica, Universidade Federal do Rio de Janeiro\\
            Rio de Janeiro, RJ 21941-972, BRAZIL}

\maketitle              

\begin{abstract}
This paper discusses work developed in recent years, in the domain of quantum optics, which has led to a better understanding of the classical limit of quantum mechanics. New techniques have been proposed, and experimentally demonstrated, for characterizing and monitoring in real time the quantum state of an electromagnetic field in a cavity.  They allow the investigation of the dynamics of the decoherence process by which a quantum-mechanical superposition of coherent states of the field becomes a statistical mixture.
\end{abstract}

\section{Introduction}
One of the most subtle problems in contemporary physics is the 
relation between the macroscopic world, described by classical physics, 
and the microscopic world, ruled by the laws of quantum mechanics. Among
the several questions involved in the quantum-classical transition,
one stands out in a striking way. As pointed out by Einstein in a letter
to Max Born in 1954~\cite{Einstein}, it concerns ``the inexistence at the
classical level of the majority of states allowed by 
quantum mechanics,'' namely coherent superpositions of classically 
distinct states. Indeed, while in the quantum world one frequently
comes across coherent superpositions of states (like in Young's two-slit 
interference experiment, in which each photon is considered to be in a
coherent superposition of two wave packets, centered around the
classical paths which stem out of each slit), one does not see
macroscopic objects in coherent superpositions of two distinct
classical states, like a stone which could be at two places at the
same time. There is an important difference between a state of this
kind and one which would involve just a classical alternative: the
existence of quantum coherence between the two localized states would
allow in principle the realization of an interference experiment,
complementary to the simple observation of the position of the object.
We know all this already from Young's experiment: the observation of
the photon path (that is, a measurement able to distinguish
through which slit the photon has passed) unavoidably destroys the
interference fringes.

If one assumes that the usual rules of quantum dynamics are valid up
to the macroscopic level, then the existence of quantum interference
at the microscopic level necessarily implies that the same phenomenon
should occur between distinguishable macroscopic states. This was
emphasized by Schr\"odinger in his famous ``cat paradox''~\cite{Schrodinger}. An important role is played by this fact in
quantum measurement theory, as pointed out by Von Neumann~\cite{Von
Neumann}.  Indeed, let us assume for instance that a microscopic
two-level system (states $|+\rangle$ and $|-\rangle$) interacts with a
macroscopic measuring apparatus, in such a way that the pointer of the
apparatus points to a different (and classically distinguishable!)
position for each of the two states, that is, the interaction
transforms the joint atom-apparatus initial state into
\begin{eqnarray*}
|+\rangle|\uparrow\rangle&\rightarrow&|+\rangle^\prime|\nearrow\,\,\rangle\,,
\\
|-\rangle|\uparrow\rangle&\rightarrow&|-\rangle^\prime|\nwarrow\,\,\rangle\,,
\end{eqnarray*}
where one has allowed for a change in the state of the two-level system, due to its interaction with the measurement apparatus.
 
The linearity of quantum mechanics implies that, if the quantum system
is prepared in a coherent superposition of the two states, say
$|\psi\rangle=(|+\rangle+|-\rangle)/\sqrt{2}$, the final state of the
complete system should be a coherent superposition of two product 
states, each of which corresponding to a different position of the
pointer:
\begin{eqnarray*}
&{}&(1/\sqrt{2})(|+\rangle+|-\rangle)|\uparrow\rangle\\
&\rightarrow&(1/\sqrt{2})(|+\rangle^\prime|\nearrow\,\,\rangle+|-\rangle^\prime|\nwarrow\,\,\rangle)=(1/\sqrt{2})(|\nearrow\,\,\rangle^\prime+|\nwarrow\,\,\rangle^\prime)\,,
\end{eqnarray*}
where in the last step it was assumed that the two-level system is 
incorporated into the measurement apparatus after their interaction
(for  instance, an atom that gets stuck to the detector). One gets,
therefore, as a result of the interaction between the microscopic and the
macroscopic system, a coherent superposition of two classically distinct states of
the macroscopic apparatus. This is actually the situation in
Schr\"odinger's cat paradox: the cat can be viewed as a measuring apparatus of the
state of a decaying atom, the state of life or death of the cat being equivalent
to the two positions of the pointer. This would imply that one should be able
in principle to get interference between the two states of the pointer:
it is precisely the lack of evidence of such phenomena in the macroscopic
world that motivated Einstein's concern. 

Faced with this problem, Von Neumann introduced through his
collapse postulate~\cite{Von Neumann} two distinct types of evolution
in quantum mechanics: the deterministic and unitary evolution
associated to the Schr\"odinger equation, which describes the
establishment of a correlation between states of the microscopic
system being measured and distinguishable classical states (for
instance, distinct positions of a pointer) of the macroscopic
measurement apparatus; and the probabilistic and irreversible process
associated with measurement, which transforms coherent superpositions of distinguishable classical states 
into statistical mixtures. This separation of the whole process into
two steps has been the object of much debate~\cite{Wigner,Wheeler,Hepp}; indeed, it would not only imply an
intrinsic limitation of quantum mechanics to deal with classical
objects, but it would also pose the problem of drawing the line
between the microscopic and the macroscopic world.

Several possibilities have been explored as solutions to this paradox,
including the proposal that a small non-linear term in the
Schr\"odinger equation, although unnoticeable for microscopic
phenomena, could eliminate the coherence between distinguishable macroscopic states,
thus transforming the quantum superpositions into statistical mixtures~\cite{Wigner}. 
The non-observability of the coherence between the two positions of the
pointer has been attributed both to the lack of non-local observables
with matrix elements between the two corresponding states~\cite{Gottfried} as well as to the fast decoherence due to
interaction with the environment~\cite{Zurek,Caldeira,Haake}. This last approach has been emphasized in recent years: decoherence follows
from the irreversible coupling of the observed system to a reservoir~\cite{Zurek,Caldeira}. In this process, the quantum
superposition is turned into a statistical mixture, for which all the
information on the system can be described in classical terms, so our
usual perception of the world is recovered.  Furthermore, for
macroscopic superpositions quantum coherence decays much faster than
the macroscopic observables of the system, its decay time being given by
the dissipation time divided by a dimensionless number measuring the
``separation'' between the two parts. The statement that these two
parts are macroscopically separated implies that this separation is an
extremely large number. Such is the case for biological systems like
``cats'' made of huge number of molecules. In the simple case
mentioned by Einstein~\cite{Einstein}, of a particle split into two
spatially separated wave packets by a distance $d$, the dimensionless
measure of the separation is $(d/\lambda_{dB})^2$, where
$\lambda_{dB}$ is the particle de Broglie wavelength. For a particle
with mass equal to 1 g at a temperature of 300 K, and $d=1$ cm, this
number is about $10^{40}$, and the decoherence is for all purposes
instantaneous. This would provide an answer to Einstein's concern:
the decoherence of macroscopic states would be too fast to be observed.

In this paper, it will be shown that the study of the interaction
between atoms and electromagnetic fields in cavities can help us
understand some aspects of this problem. In fact, many recent
contributions in the field of quantum optics have led not only to the
investigation of the subtle frontier between the quantum and the
classical world, but also of hitherto unsuspected quantum mechanical
processes like teleportation. Research on quantum optics is therefore
intimately entangled with fundamental problems of quantum mechanics.

The whole area of ``cavity quantum electrodynamics'' is a very recent
one. It concerns the interactions between atoms and discrete modes of
the electromagnetic field in a cavity, under conditions such that
losses due to dissipation and atomic spontaneous emission are very
small. Usually, one deals with atomic beams crossing cavities with a
high quality factor $Q$ (defined as the product of the angular
frequency of the mode and its lifetime, $Q=\omega\tau$). The atoms,
prepared in special states and detected after interacting with the
field, serve two purposes: they are used to manipulate the field in
the cavity, so as to produce the desired states, and also to measure
the field.

Several factors contributed to the development of this area. The
production of superconducting Niobium cavities, with extremely high
quality factors, up to the order of $10^{10}$, allows one to keep a
photon in the cavity for a time of the order of a fraction of a second. New
techniques of atomic excitation (alkaline atoms, like Rubidium and
Cesium, are frequently used for this purpose) to highly excited levels
(principal quantum numbers of the order of 50), and with maximum
angular momentum ($\ell=n-1$) --- the so-called planetary Rydberg atoms
--- have led to the production of atomic beams that interact strongly
even with very weak fields, of the order of one photon, due to the
large magnitude of the relevant electric dipoles.  Besides, the
lifetime of these states is large --- of the order of the millisecond
--- which may be understood semiclassically, from the correspondence
principle (which should be valid for $n\sim50$): the electron is
always very far away from the nucleus, and therefore its acceleration
is small, implying weak radiation and a long lifetime. One should also
mention the new techniques of atomic velocity control, which allow the
production of approximately monokinetic atomic beams, leading to a
precise control of the interaction time between atom and field. For a
review of some of the main problems and results in this field, see 
\cite{Haroche}.

\section{Coherent Superpositions of  Mesoscopic States in Cavity QED}

\subsection{Building the Coherent Superposition}

We show now how, by carefully tailoring the interactions between
two-level atoms and one mode of the electromagnetic field in a cavity,
one can produce quantum superpositions of distinguishable coherent
states of the field, thus mimicking the superposition of two
classically distinct states of a pointer.

For a harmonic oscillator, a coherent state~\cite{Glauber} is obtained by displacing the ground state in phase space. In general, the position will be displaced by $x$ and the momentum by $p$, so that the displacement can be characterized by the complex amplitude $\alpha=\sqrt{m\omega/2\hbar}(x+ip/m\omega)$, where $m$ and $\omega$ are respectively the mass and the angular frequency of the oscillator. The state is thus denoted by $|\alpha\rangle$, and it can be physically realized by applying a classical force to the oscillator. Coherent states are ``quasi-classical'' states: they evolve in time without changing their shape, the corresponding wave packet oscillating around the equilibrium position like a classical particle. Furthermore, they are minimum-uncertainty states: the product of the uncertainties in position and momentum is equal to the minimum value allowed by the Heisenberg uncertainty principle. 

A one-mode electromagnetic field can be described by a harmonic oscillator Hamiltonian, in which the position and momentum are replaced by the {\it quadratures} of the field. These are defined as the amplitudes $q_1$ and $q_2$ of the cosine and sine terms in the time-dependent expression of the field:
\be
E=E_0\left[q_1\cos(\vec k\cdot\vec r-\omega t)+q_2\sin(\vec k\cdot\vec r-\omega t)\right]\,.
\ee
This expression is analogous to the one that yields the position of a harmonic oscillator at time $t$, as a function of the initial position and momentum:
\be\label{ho}
x(t)=x(0)\cos\omega t+[p(0)/m\omega]\sin\omega t\,,
\ee
so that the quadratures $q_1$ and $q_2$ play a role analogous to the position $x(0)$ and momentum $p(0)$ (conveniently normalized). In the same way that, in quantum mechanics, the position and momentum are non-commuting operators, quantization of the electromagnetic field is achieved by requiring that the operators corresponding to the quadratures, $\hat q_1$ and $\hat q_2$, satisfy $[\hat q_1,\hat q_2]=i$. They are related to the photon annihilation operator $\hat a$ by
\begin{equation}\label{defa}
\hat a=(\hat q_1+i\hat q_2)/\sqrt{2}\,.
\end{equation}
The complex amplitude of the field is defined as $\alpha=(q_1+iq_2)/\sqrt{2}$. 

For an electromagnetic field, a coherent state also corresponds to a displaced ground state (in this case the vacuum state of the electromagnetic field). It can be explicitly written as (for a one-mode field)
\begin{equation}\label{displacement}
|\alpha\rangle=\hat D(\alpha,\alpha^*)|0\rangle=\exp\left(\alpha \ad-\alpha^*\hat a\right)|0\rangle\,,
\end{equation}
where $|0\rangle$ is the vacuum state of the electromagnetic field and $\hat D(\alpha,\alpha^*)$ is the \emph{displacement operator}. The average number of photons in the coherent state $|\alpha\rangle$ is $|\alpha|^2$~\cite{Glauber}.   This state can be physically realized by turning on a classical current (for instance, a microwave generator) when the field is in the vacuum state. The corresponding evolution operator in the interaction picture is then closely related to the displacement operator.

\begin{figure}[b]
\begin{center}
\includegraphics[width=.8\textwidth]{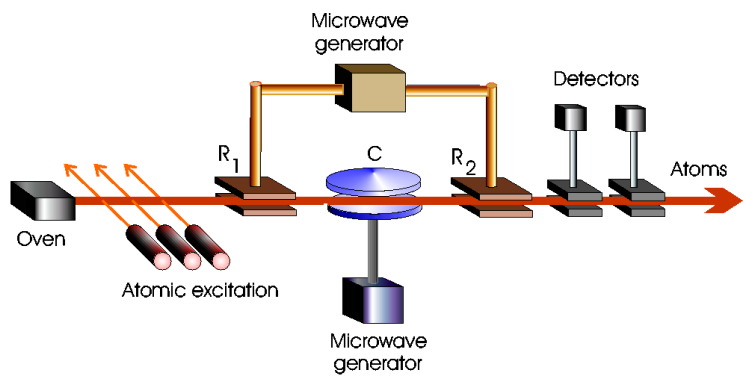}
\end{center}
\caption[]{Experimental arrangement for producing and measuring a coherent superposition of two coherent states of the field in cavity $\bf C$}
\label{cat}
\end{figure}

The method for generating the quantum superposition of two coherent states, proposed in \cite{Brune}, and sketched in Fig.~\ref{cat}, involves a beam of circular Rydberg atoms~\cite{Kleppner}
crossing a high-$Q$ cavity {\bf C} in which a coherent state is previously
injected (this is accomplished by coupling the cavity to a classical
source -- a microwave generator -- through a wave guide). The
utilization of circular levels is due to their strong coupling to
microwaves and their very long radiative decay times, which makes them
ideally suited for preparing and detecting long-lived correlations
between atom and field states~\cite{Brune2}. On either side of the
high-$Q$ cavity there are two low-$Q$ cavities ($\bf R_1$ and $\bf
R_2$), which remain coupled to a microwave generator. The fields in
these two cavities can be considered as classical. As a matter of  fact, it can be shown that, for the experiments realized so far, the average number of photons in these cavities is of the order of one. How come then this field behaves classically? This is due to the highly dissipative character of  these cavities: again, dissipation helps to turn the quantum-mechanical world into a classical one. The classical behavior of the fields in these low-$Q$ cavities was demonstrated in \cite{Ji}. 

This set of two
low-Q cavities constitutes the usual experimental arrangement in the
Ramsey method of interferometry~\cite{Brune2,Ramsey}. Two of the
(highly excited) atomic levels, which we denote by $|e\rangle$ (the upper level) and $|g\rangle$ (the lower one), are
resonant with the microwave fields in cavities $\bf R_1$ and $\bf
R_2$. The interaction of a two-level atom with a resonant electromagnetic field is analogous to  the interaction between a spin and a magnetic field: it amounts to a rotation transformation applied to the two states. The intensity of the fields in $\bf R_1$ and $\bf R_2$ is chosen so that, for the selected
atomic velocity, effectively a $\pi/2$ pulse is applied to the atom as
it crosses each cavity. For a properly chosen phase of the
microwave field, this pulse transforms the state $|e\rangle$ into the
linear combination $(|e\rangle+|g\rangle)/\sqrt2$, and the state
$|g\rangle$ into $(-|e\rangle+|g\rangle)/\sqrt2$.

Therefore, if each atom is prepared in the state $|e\rangle$ just
prior to crossing the system, after leaving $\bf R_1$ the atom is in a
superposition of two circular Rydberg states $|e\rangle$ and
$|g\rangle$:
\begin{equation}
|\psi_{\rm atom}\rangle={1\over\sqrt2}(|e\rangle+|g\rangle)\,.\label{atom}
\end{equation}

\begin{figure}[b]
\begin{center}
\includegraphics[width=.3\textwidth]{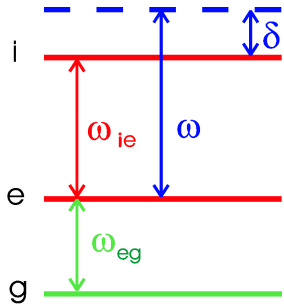}
\end{center}
\caption[]{Atomic level scheme: The transition $i\leftrightarrow e$ is detuned
  by $\delta $ from the frequency $\omega $ of a mode of cavity ${\bf
    C}$,
while the transition $e\leftrightarrow g$ is resonant with the fields in $%
{\bf R_1}$ and ${\bf R_2}$. State $|g\rangle $ is not affected by the field
in ${\bf C}$}
\label{f2}
\end{figure}

On the other hand, the superconducting cavity is assumed not to be in
resonance with any of the transitions originating from those two
atomic states. This means that the atom does not suffer a transition,
and does not emit or absorb photons from the field. This property is
further enhanced by the fact that the cavity mode is such that the
field slowly rises and decreases along the atomic trajectory, so that,
for sufficiently slow atoms, the atom-field coupling is adiabatic.
However, the cavity is tuned in such a way that it is much closer to
resonance with respect to one of those transitions, say the one
connecting $|e\rangle$ to some intermediate state $|i\rangle$. The
relevant level scheme is illustrated in Fig.~\ref{f2}. This implies
that, if the atom crosses the cavity in state $|e\rangle$, dispersive
effects can induce an appreciable phase shift on the field in the
cavity. That is, the atom acts like a refraction index, changing the frequency of the field while the interaction is on --- the corresponding energy change is just the AC-Stark shift, which for a Fock state of the electromagnetic field is proportional to the number of photons in the cavity. This frequency shift, multiplied by the interaction time between the atom and the mode, leads to a phase shift of the field in the cavity, if the atom is in state $|e\rangle$.  The phase shift is negligible, however, if the atom is in
state $|g\rangle$.  For a principal quantum number equal to 50 in the
state $e$, and the cavity tuned close to the $50\rightarrow51$
circular to circular transition (around 50 GHz), a phase shift of the
order of $\pi$ per photon is produced by an atom crossing the centimeter size
cavity with a velocity of about 100 m/s~\cite{Brune}.

Note that, if there is a coherent state in the cavity, a phase shift of $\phi$ per photon if the atom is in state $|e\rangle$ implies that
\begin{equation}
|e\rangle|\alpha\rangle\rightarrow |e\rangle \exp(-|\alpha|^2/2)\sum_{n=0}^\infty {\alpha^n\over \sqrt{n!}}e^{in\phi}|n\rangle=|e\rangle|\alpha e^{i\phi}\rangle\,,
\end{equation}
that is, the phase of the coherent state is shifted by $\phi$. The above equation makes use of the expansion of a coherent state in terms of Fock states. 

After the atom has crossed the cavity, in a time short compared to the
field relaxation time and also to the atomic radiative damping time,
the state of the combined atom-field system can be written as
\begin{equation} |\psi_{\rm
atom+field}\rangle={1\over\sqrt2}(|e;-\alpha\rangle+
|g;\alpha\rangle)\,,\label{entangled1}
\end{equation}
assuming that the phase shift is $\pi$ per photon if the atom is in the excited
state. The entanglement between the field and atomic states is
analogous to the correlated two-particle states in the
Einstein-Podolski-Rosen (EPR) paradox.\cite{EPR,Bell,Aspect} The two
possible atomic states $e$ and $g$ are here correlated to the two
field states $|-\alpha\rangle$ and $|\alpha\rangle$, respectively.
After the atoms leave the superconducting cavity, one can detect them
in the $e$ or $g$ states, by sending them through two ionization
chambers, the first one having a field smaller than the second, so
that it ionizes the atom in the $e$ state, but not in the $g$ state,
while the second ionizes the atoms that remain in state $g$ (Fig.~
\ref{cat}). In the actual experiment, this detection system is replaced by a single chamber, with a static electric field that increases linearly along the direction of atomic motion. This measurement projects the field in the cavity either
onto the state $|\alpha\rangle$ (if the atom is detected in state $g$), or onto
the state $|-\alpha\rangle$ (if the atom is detected in state $e$).
However, as in an EPR experiment~\cite{Aspect}, one may choose to make
another kind of measurement, letting the atom cross, after it leaves
the superconducting cavity, a second classical microwave field ($\bf
R_2$ in Fig.~\ref{cat}), which amounts to applying to the atom
another $\pi/2$ pulse.  The state (\ref{entangled1}) gets transformed
then into
\begin{equation}
|\psi^{\prime}_{\rm
atom+field}\rangle=\hbox{${1\over2}$}\left(|e;-\alpha\rangle-|e;\alpha\rangle+
|g;\alpha\rangle+|g;-\alpha\rangle\right)\,.\label{entangled2}
\end{equation}
If one detects now the atom in the state $|g\rangle$ or $|e\rangle$, the
field is projected onto the state
\begin{equation}
|\psi_{\rm
cat}\rangle={1\over N_1} \left(|\alpha\rangle+e^{i\psi_1}
|-\alpha\rangle\right)\,,\label{cat1}
\end{equation}
where $N_1=\sqrt{2\left[1+\cos\psi_1\exp(-2|\alpha|^2|) \right]}$ and
$\psi_1=0$ or $\pi$, according to whether the detected state is $g$ or
$e$, respectively. One produces therefore a coherent superposition of
two coherent states, with phases differing by $\pi$. For
$|\alpha|^2\gg1$, this is a ``Schr\"odinger cat-like'' state.

Superpositions of coherent states of the field were produced in the
experiment reported in \cite{Brune3}, and were detected by a
procedure proposed in \cite{Luiz1,Luiz2}. 

\subsection{Measuring the Coherent Superposition}

Once the quantum superposition is produced, how could one tell the difference between such a superposition and a statistical mixture of the two coherent states? This can be done by simply sending another atom, in the same initial state as the first one. It can be shown then~\cite{Luiz2} that, for the state (\ref{cat1}), with $|\alpha|\gg1$, there is a perfect correlation between the measurements of the first and the second atom: both are always detected in the same state. On the other hand, for the corresponding statistical mixture the probability of detecting the second atom in state $|e\rangle$ is 50\%, independently of which state was detected for the first atom. By delaying the sending of the second atom, one may thus explore the dynamical process by which the quantum superposition is transformed into a statistical  mixture, due to the always present dissipation  in a non-perfect cavity. 

\begin{figure}[b]
\begin{center}
\includegraphics[width=.5\textwidth]{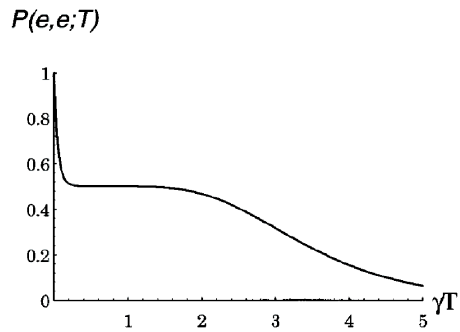}
\end{center}
\caption[]{Conditional probability for finding the second atom in state $|e\rangle$ if the first atom was detected in state $|e\rangle$, as a function of time (measured in units of the field damping time)}\label{dec}
\end{figure}

The time-dependent behavior of the conditional probability for measuring the second atom in the upper state, knowing that the first atom was also measured in the upper state, is displayed in Fig.~\ref{dec}. The sharp decay of this conditional probability from the perfectly coherent situation to the plateau associated with an incoherent superposition defines the {\it decoherence time}. This time can be shown to be equal to the dissipation time for the field in the cavity divided by twice the average number of photons in the field. Thus, it becomes shorter as the field becomes more macroscopic. Note also that the plateau eventually disappears, and the probability for measuring the second atom in the state $|e\rangle$ goes to zero. This can be easily understood: the field in the cavity {\bf C} leaks out, and therefore the sole effect on the atom initially prepared in the state $|e\rangle$ is the sum of two $\pi/2$ pulses in the cavities $\bf R_1$ and $\bf R_2$, that is a $\pi$ pulse, which takes the atom into the state $|g\rangle$.

An experimental realization of this proposal was made in 1996 by Haroche's group at Ecole Normale Sup\'erieure, in Paris~\cite{Brune3}. The dynamical measurement of the decoherence process, as proposed above, was in agreement with the theoretical predictions.

\section{The Wigner Distribution}

One might wonder if it could be possible to get, from the above experimental setup, a more complete information on the field in the cavity. This was shown to be indeed possible in \cite{Lutter}: a slight modification of the above experiment leads to the reconstruction of the so-called Wigner distribution of the field in the cavity, which provides a complete description of the quantum state of the field in phase space.

Phase space probability distributions are very useful in classical statistical physics. Averages of relevant functions of the positions and momenta of the particles can be obtained by integrating these functions with those probability weights. 

In quantum mechanics, similar averages are calculated by taking the trace of the product of the density operator that describes the system with the observable of interest. Heisenberg's inequality forbids the existence in phase
space of bonafied probability distributions, since one cannot
determine simultaneously the position and the momentum of a particle.
In spite of this, phase space distributions may still play a useful
role in quantum mechanics, allowing the calculation of the average of
operator-valued functions of the position and momentum operators as
classical-like integrals of $c$-number functions. These functions are
associated to those operators through correspondence rules, which
depend on a previously defined operator ordering.

From all phase space representations, the Wigner distribution~\cite{Wigner2} is the most natural one, when one looks for a quantum-mechanical analog of a classical probability distribution in phase space. It is in fact the only distribution that leads to the correct marginal distributions, for any direction of integration in phase space~\cite{Bertrand,Ulf}. Let us consider for simplicity a one-dimensional problem, for a particle with position $q$ and momentum $p$. We take these to be dimensionless variables, measured in terms of some typical position and momentum of the system, which play the role of natural units (for a harmonic oscillator, the natural units would be the uncertainties in position and momentum of the ground state). If the state of the particle is characterized by the density operator $\hr$, then we should have not only 
\begin{equation}\label{pq}
\int \D p\,W(q,p)=\langle q|\hr|q\rangle\,,\int \D q W(q,p)=\langle p|\hr|p\rangle\,,
\end{equation}
where $|q\rangle$ and $|p\rangle$ are eigenstates of the operators $\hat q$ and $\hat p$, respectively, but also
\begin{equation}\label{ptheta}
P(q_\theta)=\int W(q_\theta\cos\theta-p_\theta\,{\sin}\,\theta, q_\theta\,{\sin}\,\theta+p_\theta\cos\theta)\D p_\theta\,.
\end{equation}
where now
\begin{equation}\label{pquantum}
P(q_\theta)=\langle q_\theta|\hr|q_\theta\rangle\,,
\end{equation}
the rotated coordinate $q_\theta$ being defined as
\be
q_\theta=q\cos\theta+p\sin\theta\,.
\ee

One should note that, for a pure state, $\langle q|\hr|q\rangle=|\psi(q)|^2,\, \langle p|\hr|p\rangle=|\tilde\psi(p)|^2$. One should also note that from (\ref{pq}) it follows immediately the normalization property:
\begin{equation}\label{norm Wigner}
\int \D p\, \D q\,W(q,p)=1\,.
\end{equation}

Expression (\ref{ptheta}), which yields the probability distribution
for $q_\theta$ in terms of the function $W(q,p)$, is called a {\it
 Radon transform}. Note that this transform may be defined independently of quantum mechanics, and in fact it was investigated in 1917 by the mathematician
Johan Radon~\cite{Radon}. He showed that, if one knows $P(q_\theta)$
for all angles $\theta$, then one can uniquely recover the function $W(q,p)$,
through the so-called {\it Radon inverse transform}. Quantum mechanics comes into play if one now
identifies $P(q_\theta)$, given by the Radon transform (\ref{ptheta}),
with the quantum expression (\ref{pquantum}). It follows then that
(\ref{ptheta}) and (\ref{pquantum}) uniquely determine the function
$W(q,p)$, in terms of the density operator $\hr$ of the system. The
function $W(q,p)$ is in this case precisely the Wigner function of the
system, expressed in terms of the density matrix in the position representation by
\begin{equation}\label{w}
W(q,p)={1\over2\pi}\int_{-\infty}^{+\infty}e^{ipx}\left\langle q-{x\over2}\left|\hr\right|q+{x\over2}\right\rangle\,\D x\,,
\end{equation}
which, except for a normalization constant, is the famous expression
written down by Wigner \cite{Wigner2} in his article {\it ``On the
Quantum Correction for Thermodynamic Equilibrium,''} published in
1932.

\begin{figure}[b]
\begin{center}
\includegraphics[width=.4\textwidth]{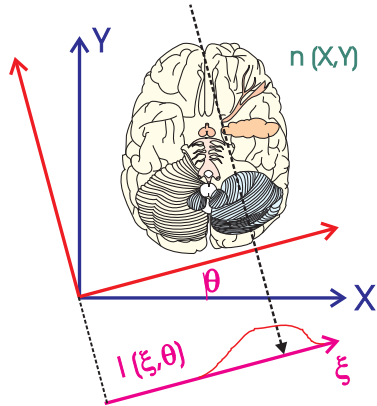}
\end{center}
\caption[]{Medical tomography: Measurement of the X-ray absorption for all angles along a plane allows one to reconstruct the absorptive part of the refraction index for a slice of the organ under investigation}\label{tomo}
\end{figure}

The demonstration of this result can be found in \cite{Bertrand,Ulf}. Let us note that Radon's result is
the mathematical basis of tomography. In fact, application of this
procedure to medicine (see Fig.~\ref{tomo}) has brought the Nobel
prize in Medicine to Cormack and Hounsfield in
1979.

The tomographic procedure has a simple interpretation for a harmonic
oscillator. From (\ref{ho}), it is clear that in this case measuring
the quadratures for all angles is equivalent to measuring the position
of the harmonic oscillator for all times from $0$ to $2\pi/\omega$.
This implies that the measurement of $|\psi(x,t)|^2$ for
$0<t\le2\pi/\omega$ allows one to reconstruct the state $\psi(x,t)$ of
the harmonic oscillator.

The question about what is the minimum set of measurements needed to reconstruct the state of a system is actually a very old problem in quantum mechanics. In his article on quantum mechanics in the {\it Handbuch der Physik} in 1933 \cite{Pauli}, Pauli stated that ``the mathematical problem, as to whether for given functions $W(x)$ and $\tilde W(p)$ [probability distributions in position and momentum  space], the wave function $\psi$, if such a function exists, is always uniquely determined has still not been investigated in all its generality.''  One knows now the answer to this question: the probability distributions $W(x)$ and $\tilde W(p)$ do not form a complete set in the tomographic sense, and therefore are not sufficient to determine uniquely the quantum state of the system. 

In 1949, it was shown by Moyal~\cite{Moyal} that the Wigner distribution can be
used to calculate averages of symmetric operator functions of $q$ and
$p$, as classical-like integrals in phase space. Thus, for instance,
\begin{equation}
{\rm Tr}\left(\hr\left\{\hat q^2\hat p\right\}_{\rm sim}\right)={\rm Tr}\left[\hr\left(\hat q^2\hat p+\hat q\hat p\hat q +\hat p\hat q^2\right)/3\right]=\int \D q\D p\,W(q,p)q^2p\,,
\end{equation}
where $W(q,p)$ is the Wigner function corresponding to the density operator $\hr$. 

For a harmonic oscillator, an alternative expression for the Wigner function may be obtained by expressing the position and momentum operators $\hat q$ and $\hat p$ (or, alternatively, the quadrature operators $\hat q_1$ and $\hat q_2$) in terms of the annihilation and creation operators $\hat a$ and $\hat a^\dagger$, defined by (\ref{defa}).

One gets then \cite{Cahill}:
\begin{equation}\label{wa}
W(\alpha,\alpha^*)=2{\rm Tr}\left[\hr \hat D(\alpha,\alpha^*)e^{i\pi\ad\hat a}\hat D^{-1}(\alpha,\alpha^*)\right]\,,
\end{equation}
where the displacement operator is defined by (\ref{displacement}). Since $\hat{\cal P}=\exp(i\pi \hat a^\dagger \hat a)$ is the parity operator (note that $\hat{\cal P}\hat q\hat{\cal P}=-\hat q$, $\hat{\cal P}\hat p\hat{\cal P}=-\hat p$), this expression shows that the Wigner function is proportional to the average of the displaced parity operator. 

The Wigner function given by (\ref{wa}) involves actually a different normalization with respect to the one defined by (\ref{w}): one must set $W\rightarrow 2\pi W$, so that 
\be\label{wan}
\int (\D^2\alpha/\pi)\, W(\alpha,\alpha^*)=1\,.
\ee

It is easy to check that the Wigner function is real and bounded. With the normalization (\ref{wan}), it satisfies the bound 
\begin{equation}\label{bound}
|W((\alpha,\alpha^*)|\le2\,.
\end{equation}
However, it may become negative: this is related to the fact that a bonafied phase space distribution cannot exist in quantum mechanics. 

\subsection{Measuring the Wigner Function}

It was only in 1989 that Risken and Vogel suggested that the technique of homodyne detection could be used to reconstruct the Wigner function of a running electromagnetic wave~\cite{Vogel}. Indeed, this technique allows the measurement of the probability distribution of an arbitrary quadrature of the electromagnetic field $q_\theta=q_1\cos\theta+q_2\sin\theta$, and   one is then able to reconstruct the Wigner function through the inverse Radon transform.

The first experimental demonstration of this procedure was achieved in 1993 by Smithey {\it et al}~\cite{Smithey}. 
In view of the low detection efficiency in those experiments, the
detected distribution was actually a smoothed version of the Wigner
function, closely related to the so-called Husimi distribution. A much better
result was achieved by Mlynek's group in 1995~\cite{Mlynek}, clearly
displaying a highly compressed Gaussian, corresponding to the
experimentally obtained Wigner function of a squeezed state of light
emerging from an optical parametric oscillator (squeezed states are minimum uncertainty states such that the variance of one of the quadratures is smaller than the one corresponding to the vacuum state of the field). A procedure closely
related to the homodyne detection method was used to reconstruct the
vibrational state of a molecule by T. J. Dunn \emph{et al}~\cite{Dunn}.

Using a different (but also indirect) method, the Wigner function of the center-of-mass state of an ion trapped in a harmonic trap, and placed in the first excited state of the harmonic potential, was measured by Wineland's group at NIST~\cite{Leibfried}.

\begin{figure}[b]
\begin{center}
\includegraphics[width=.8\textwidth]{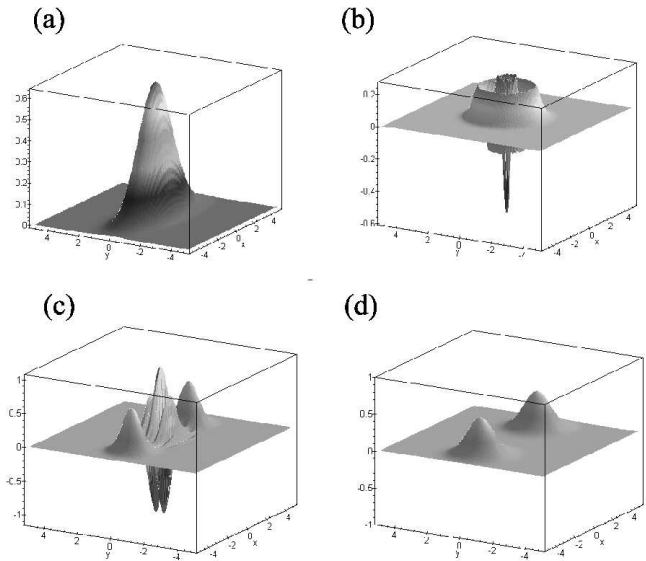}
\end{center}
\caption[]{Examples of Wigner distributions: (a) Squeezed state; (b) Harmonic oscillator eigenstate with $n=3$; (c) Superposition of two coherent states, $|\psi\rangle\propto|\alpha_0\rangle+|-\alpha_0\rangle$, with $\alpha_0=3$; (d) Statistical mixture $\hbox{${1\over2}$}(|\alpha_0\rangle\langle\alpha_0|+|-\alpha_0\rangle\langle-\alpha_0|)$, also with $\alpha_0=3$}
\label{f02}
\end{figure}

\subsection{Examples of  Wigner Functions}

Some examples of Wigner functions are shown in Fig.~\ref{f02}. The Wigner function corresponding to the ground state of a harmonic oscillator (or the vacuum of the electromagnetic field) is a Gaussian, centered around the origin of phase space. For a squeezed state, one gets a compressed Gaussian. On the other hand, for eigenstates of the harmonic oscillator -- corresponding, for an electromagnetic field, to states with well-defined number of  photons -- the Wigner function is negative in some regions of phase space, as shown in Fig.~\ref{f02}(b). As mentioned before, this is an evidence that it cannot be considered a bonafied probability distribution. Note also that, while the statistical mixture of two coherent states (which are displaced ground states) corresponds to a sum of two Gaussians, the Wigner function corresponding to the quantum superposition of two coherent states exhibits interference fringes, a clear signature of coherence. Decoherence leads to the disappearance of these fringes. Therefore, the measurement of the Wigner function of the
electromagnetic field would be a clear-cut way of distinguishing
between a coherent superposition and a mixture of the two coherent
states. Furthermore, if one could make this measurement fast enough, one would be able to follow the decoherence process in real time.

\section{Direct Measurement of the Wigner Function}

Once the proper state of the field is produced in the cavity, how
would one be able to measure it? As shown in \cite{Lutter,Lutter2}, it is
actually possible to measure the Wigner function of the field by a
relatively simple scheme, which provides directly the value of the
Wigner function at any point of phase space. This is in contrast with
the tomographic procedure, or the method adopted at NIST, which yield the Wigner function only
after some integration or summation. Furthermore, and also in contrast
with those methods, the scheme proposed in \cite{Lutter,Lutter2} is not sensitive to detection
efficiency, as long as one atom is detected within a time shorter than
the decoherence time. A similar procedure can be applied to the
reconstruction of the vibrational state of a trapped ion~\cite{Lutter}, and also in some cases to molecules~\cite{molecule}. We will discuss here only the application to the electromagnetic field.

The basic experimental scheme for measuring the Wigner function~\cite{Lutter} coincides with the one used to produce the ``Schr\"odinger cat''-like state, illustrated in Fig.~\ref
{cat}. A high-$Q$ superconducting cavity ${\bf C}$ is placed between two low-$%
Q$ cavities (${\bf R_1}$ and ${\bf R_2}$ in Fig.~\ref{cat}). The cavities $%
{\bf R_1}$ and ${\bf R_2}$ are connected to the same microwave
generator.  Another microwave source is
connected to ${\bf C}$, allowing the injection of a coherent state into 
this cavity.  This system is
crossed by a velocity-selected atomic beam, such that an atomic
transition $e\leftrightarrow g$ is resonant with the fields in ${\bf
  R_1}$ and ${\bf R_2}$, while another transition $e\leftrightarrow i$
is quasi-resonant (detuning $\delta $) with the field in ${\bf C}$, so
that the atom interacts dispersively with this field if it is in state
$e$, while no interaction takes place in $\bf C$ if the atom is in
state $g$. The relevant level scheme is shown in Fig.~\ref{f2}.  Just
before ${\bf R_1}$,
the atoms are promoted to the highly excited circular Rydberg state $%
|e\rangle $ (typical principal quantum numbers of the order of 50,
corresponding to lifetimes of the order of some milliseconds). As each atom
crosses the low-$Q$ cavities, it sees a $\pi /2$ pulse, so that
$||e\rangle \rightarrow [|e\rangle +|g\rangle ]/\sqrt{2}$, and
$|g\rangle \rightarrow [-|e\rangle +|g\rangle ]/\sqrt{2}$.  If the atom is in state $e$ when crossing $\bf
C$, there is an energy shift of the atom-field system (Stark shift), which
dephases the field, after an effective interaction time between the atom and the cavity mode. We assume that the one-photon phase shift is equal to $\pi$. We call this a conditional phase shift, since it depends on the atomic state. 

The atom is detected and the experiment is repeated many times, for each
amplitude and phase of the injected field $\alpha$, starting from the same
initial state of the field. In this way, the probabilities $P_e$ and $%
P_g$ of detecting the probe atom in states $e$ or $g$ are determined. It was shown in \cite{Lutter2} that
\begin{equation}\label{pegp}
P_g-P_e=W(-\alpha,-\alpha^{*})/2\,,  \label{Delta P'}
\end{equation}
where the Wigner function in this expression is defined in (\ref{wa}), with the normalization (\ref{wan}). Therefore, the difference between the two probabilities yields a direct
measurement of the Wigner function!

The derivation of (\ref{Delta P'}), developed in \cite{Lutter,Lutter2}, is based on expression (\ref{wa}) for the Wigner function. Indeed, one may notice that the experimental procedure discussed above amounts to implementing experimentally on the state to be reconstructed the two operations explicitly represented in (\ref{wa}): the displacement operation (implemented through the injection of the coherent microwave field) and the parity operation (implemented through the conditional $\pi$-phase shift). In particular, the distribution in (\ref{Delta P'}) clearly satisfies (\ref{bound}), since $|P_g-P_e|\le2$. 

An important feature of this scheme is the insensitivity to the
detection efficiency of the atomic counters, of the order of
40$\pm$15\% in recent experiments~\cite{Brune}. 

One should note that this method allows the measurement of the Wigner
function at each time $t$, allowing therefore the monitoring of the
decoherence process ``in real time.''  It is interesting, in this
respect, to compare the procedure described above with the one described before in this article, as 
proposed in \cite{Luiz2}, with the objective of
observing the decoherence of a Schr\"{o}dinger cat-like state. As we have seen, it was proposed in that reference that the decoherence of the state $|\pm
\rangle =\left( |\alpha \rangle \pm |-\alpha \rangle \right) /N_{\pm
  }$ could be observed by measuring the joint probability of detecting
in states $|e\rangle$ or $|g\rangle$ a pair of atoms, sent through the
system depicted in Fig.~\ref{cat}, both atoms being prepared initially in the same state. Detection of the first atom prepares the coherent
superposition of coherent states. Detection of the
second atom probes the state produced in $\bf C$.  Since no field was
injected into the cavity between the two atoms, it is clear now that
the experiment proposed in \cite{Luiz2} amounts to a measurement
of the Wigner function at the origin of phase space, which is non zero for the pure
state $|\pm \rangle $, vanishes after the decoherence time, and increases again as dissipation takes place, bringing the
field to the vacuum state. In the experiment realized by Brune {\it et al}~\cite{Brune}, both $|e\rangle $ and
$|g\rangle $ lead to dephasings (in opposite directions) of the field in $%
{\bf C}$.  In this case, it is easy to show that the
Wigner function is again recovered, as long as the one-photon phase shift
is $\phi =\pi /2$ (with opposite signs for $e$ and $g$), and a
dephasing $\eta =\pi /2$ is applied to the second Ramsey zone~\cite{Lutter}.

Getting a $\pi$ phase shift per photon imposes stringent conditions on the experiment. The interaction time between the atom and the cavity field should be large enough, which implies using slow atoms, with a precisely controlled speed. Furthermore, the interaction time between the atom and the cavity field should be much smaller than the damping time of the field in the cavity, and therefore a very good cavity is required.

\begin{figure}[b]
\begin{center}
\includegraphics[width=\textwidth]{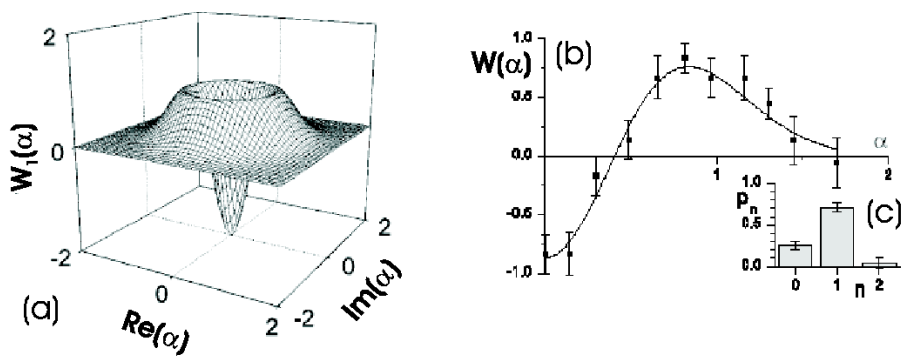}
\end{center}
\caption[]{Wigner function for a one-photon state: (a) Distribution in phase space; (b) Distribution measured in \cite{Wigner1}; (c) Corresponding photon-number distribution, showing that one does not have a pure one-photon state,  due to imperfections in the preparation process, the possible decay of the photon in the cavity, and the contamination with thermal photons}
\label{W2}
\end{figure}

An easier task consists in measuring the value of the Wigner function at the origin of phase space when one knows beforehand that the field in the cavity contains at most one photon. In this case, one does not need to inject a field into the cavity, and the dispersive interaction leading to the phase shift of the field can be replaced by a resonant $2\pi$ interaction between levels $e$ and $i$ (see Fig.~\ref{f2}). This interaction takes the atom from state $e$ to state $i$ and then back to state $e$ (thus the name ``$2\pi$-rotation'', in view of the analogy with the full rotation of a spin 1/2), if there is one photon in the field. Exactly as it would happen with a spin 1/2 object, the state changes sign under this transformation: $|e\rangle|1\rangle_{\rm field}\rightarrow -|e\rangle|1\rangle_{\rm field}$. On the other hand, nothing happens if the atom is in state $g$ or if there is no photon in the field. The conditional one-photon $\pi$ phase change is thus accomplished in this case with a resonant interaction, which requires an interaction time much shorter than the dispersive case. This idea was implemented in an experiment at Ecole Normale Sup\'erieure, in Paris~\cite{Wigner0}.  The one-photon state was produced by sending an excited atom through the empty cavity, where the atom suffers a $\pi$ transition, leaving one photon in the cavity, from which it exits in the state $g$. This was the first time a negative value was measured for the Wigner function of an electromagnetic field, namely the value at the origin of the Wigner function corresponding to a one-photon state [this distribution is shown in Fig.~\ref{W2}(a)].

More recently~\cite{Wigner1}, the Paris group was able to measure the full Wigner function for a one-photon state in the cavity, using the technique proposed in \cite{Lutter}. The result is displayed in Fig.~\ref{W2}(b), which exhibits a slice of the cylindrically symmetric distribution. From the Wigner function, it is possible to get the photon-number distribution, which is displayed in Fig.~\ref{W2}(c). This distribution shows that the state produced in the cavity is not a perfect one-photon state, which explains the fact that the value of the Wigner function at the origin of phase space is larger than $-2$, the value it should have for a one-photon state. An interesting feature of this measurement is that it probes a region of the phase space with area smaller than $\hbar$, which corresponds to the negative region of the Wigner function displayed in Fig.~\ref{W2}. It is thus an explicit demonstration of the fact that it is possible in principle to probe regions of phase space as small as one wants!

\section{Conclusion}

Since the invention of the laser, the field of quantum optics has been a very active field. Its discoveries have had not only an important technological impact, but have also led to experiments and proposals that probe fundamental questions of quantum mechanics. Some of these questions were discussed here: experiments in the field of cavity quantum electrodynamics have helped us to probe the subtle boundary between the classical and the quantum world, and have allowed the monitoring of the decoherence process, which is at the heart of quantum theory of measurement. The development of new techniques for probing the quantum state of the electromagnetic field in a cavity have led to the experimental unveiling of the Wigner function of a one-photon field, thus demonstrating the feasibility of probing regions of phase space with area smaller than $\hbar$. These methods may lead to a new generation of experiments, which will probe the dynamics of the quantum state of the electromagnetic field.

New challenges involve the demonstration of the teleportation process between two-level atoms~\cite{telep}, as well as trying to control the decoherence process, which is the main villain of quantum computers. Several proposals for fighting decoherence have been made in the last years, ranging from quantum error correction schemes~\cite{Shor} to feedback implementations~\cite{Mabuchi,Milburn}, from the realization of $q$-bits in symmetric subspaces
decoupled from the environment~\cite{Zanardi} to dynamical
decoupling techniques~\cite{Viola} and reservoir engineering~\cite{Luiz4}.

On a fundamental level, difficult problems still persist, related to the classical
limit of non-linear systems, where chaotic behavior may play an
important role~\cite{Zurek2}. 

 Even though fundamental problems related to the classical limit of quantum
mechanics and the quantum theory of measurement remain to be solved, I
think it is fair to say that quantum optics has helped us to
understand and observe an important piece of this puzzle.

\section*{Acknowledgments}

This work was partially supported by PRONEX (Programa de Apoio a
N\'ucleos de Excel\^encia), CNPq (Conselho Nacional de Desenvolvimento
Cient\'\i fico e Tecnol\'ogico), FAPERJ (Funda\c c\~ao de Amparo \`a Pesquisa do Estado do Rio de Janeiro), FUJB (Funda\c c\~ao Universit\'aria
Jos\'e Bonif\'acio), and the Millennium Institute on Quantum Information.  It is a pleasure to  acknowledge the collaboration on 
the subjects covered by this paper with my students A. R. R. Carvalho,  M. Fran\c ca Santos, T.B.L. Kist, L.G. Lutterbach, and P. Milman, and with my colleagues M. Brune, S. Haroche, R.L. de Matos Filho, M. Orszag, J.M. Raimond, and N. Zagury.

\end{document}